\documentclass[aps,prb,twocolumn,superscriptaddress, show pacs]{revtex4-2}

\usepackage{graphicx}
\usepackage{dcolumn}
\usepackage{bm}
\usepackage{hyperref}
\usepackage{bm, amsmath}
\begin{document}

\title{Low-Temperature Skyrmions and Spiral Reorientation Processes in Chiral Magnets with Cubic Anisotropy: Guidelines for Bridging Theory and Experiment}

\author{A. O. Leonov}
\thanks{leonov@hiroshima-u.ac.jp}
\affiliation{Department of Chemistry, Faculty of Science, Hiroshima University Kagamiyama, Higashi Hiroshima, Hiroshima 739-8526, Japan}
\affiliation{International Institute for Sustainability with Knotted Chiral Meta Matter (WPI-SKCM2), Hiroshima University, 1-3-1 Kagamiyama, Higashi-Hiroshima, Hiroshima 739-8526, Japan}
\author{G. G\"odecke}
\thanks{g.goedecke@tu-braunschweig.de}
\affiliation{Institut f\"ur Physik der Kondensierten Materie, TU Braunschweig, D-38106 Braunschweig, Germany}
\author{J. Grefe}
\affiliation{Institut f\"ur Physik der Kondensierten Materie, TU Braunschweig, D-38106 Braunschweig, Germany}
\author{S. S\"ullow}
\affiliation{Institut f\"ur Physik der Kondensierten Materie, TU Braunschweig, D-38106 Braunschweig, Germany}
\author{D. Menzel}
\affiliation{Institut f\"ur Physik der Kondensierten Materie, TU Braunschweig, D-38106 Braunschweig, Germany}

\date{\today}

\begin{abstract}
We revisit the phenomenological Dzyaloshinskii framework, a central theoretical approach for describing magnetization processes in bulk chiral magnets, and demonstrate how magnetocrystalline cubic anisotropy reshapes the phase diagrams of states and provides the key mechanism stabilizing low-temperature skyrmion phases.
We show that, for magnetic field directions along the easy anisotropy axes, the phase diagrams feature stable skyrmion pockets for both signs of the anisotropy constant. 
We further analyze the nature of the transitions at the critical field $H_{c1}$, associated with the reorientation of stable and metastable spirals along the field. We also examine the transition at $H_{c2}$, where the conical state closes into the homogeneous state accompanied by a deviation of the wave vector from the field direction.
By mapping characteristic anisotropy-dependent parameters in the theoretical phase diagrams, we provide guidelines for connecting theory with experiment and for estimating the cubic anisotropy constant in Fe$_{1-x}$Co$_x$Si and MnSi. Our results indicate that samples Fe$_{1-x}$Co$_x$Si with small $x \sim 0.1$ possess sufficiently strong cubic anisotropy to stabilize a low-temperature skyrmion phase. 
Overall, these theoretical findings establish a quantitative framework for predicting and interpreting skyrmion stability in other cubic helimagnets as well.
\end{abstract}

\maketitle

\section{Introduction}

Bulk cubic helimagnets such as MnSi \cite{Ishikawa,Muehlbauer09}, FeGe \cite{Ludgren70,Wilhelm2011}, and Cu$_2$OSeO$_3$ \cite{Seki2012,Adams2012} represent archetypal systems where chiral magnetic skyrmions \cite{JETP89,Bogdanov94}—two-dimensional whirls of magnetization—were first discovered. This breakthrough sparked intense research of the distinctive topological and magnetic properties of skyrmions \cite{yuFeCoSi,yuFeGe}, as well as their potential applications in next-generation information storage and data processing technologies \cite{Tomasello14,Cortes-Ortuno}.

In bulk chiral magnets, the magnetic behavior is governed by a well-defined hierarchy of competing interactions \cite{Bak80,Nakanishi80,Belitz2006}: the dominant isotropic exchange, the intermediate-strength Dzyaloshinskii–Moriya interaction (DMI), and comparatively weak magnetic anisotropies \cite{Butenko10,cubic}.
While the general features and solutions of chiral modulated phases in all chiral magnets are primarily determined by the isotropic energy functional, additional weaker interactions 
can subtly reshape the energy landscape and favor the stabilization of selected phases. Most notably, these effects promote the formation of the hexagonal skyrmion lattice (SkL) within the A-phase pocket near the ordering temperature $T_c$ \cite{Kadowaki82}. In this high-temperature region, skyrmion stability is enhanced by thermal fluctuations \cite{Muehlbauer09,Buhrandt2013,Janoschek}, the softening of the magnetization modulus \cite{Wilhelm2011,leonov2018}, and dipolar interactions.
In general, the stability of the "high-temperature" skyrmion lattice (HT-SkL) is governed by the energy difference between the SkL, $W_{\rm SkL}$, and the conical spiral, $W_{\rm cone}$, as calculated within the isotropic Dzyaloshinskii model (Eq.~\ref{density}). Defining $\Delta W_{\rm min} = W_{\rm SkL} - W_{\rm cone}$, one finds that its minima trace a curve $\xi(T)$ corresponding precisely to the magnetic-field range within the A-phase \cite{Roessler,cubic,leonov2018}. 
As a result, cubic helimagnets are established to host a common set of magnetic states---one-dimensional helical and conical spirals, along with two-dimensional skyrmion crystals---that together form a universal magnetic phase diagram \cite{Grigoriev2006,Grigoriev2006_2,Bauer}.

A similar reasoning applies to the stability of magnetic phases far from $T_c$, where the dominant influence arises from magnetic anisotropies \cite{droplets}. The most immediate effect of the cubic anisotropy is to lock the propagation directions of spiral states along the easy anisotropy axes below the critical field $H_{c1}$, often leading to the formation of multidomain spiral states \cite{Seki2012,Bak80,Plumer}. Furthermore, the cubic anisotropy determines the nature of the transition between the conical and homogeneous states at the critical field $H_{c2}$ \cite{cubic,droplets,Baral}. 

A sufficiently strong cubic anisotropy has also been shown to render the thermodynamic stability of the low-temperature skyrmion phase (LT-SkL), thereby causing deviations from the universal behavior.
Among B20 cubic helimagnets, Cu$_2$OSeO$_3$ is currently the only known example  exhibiting a LT-SkL pocket  that emerges near zero temperature \cite{Bannenberg2019,Halder}. The theoretical explanation for the anisotropy-induced stabilization of the SkL primarily relies on its effect on one-dimensional conical states—the main competitor of skyrmions \cite{Chacon2018,Bannenberg2019}. Specifically, the ideal magnetization rotation in cones can be controllably distorted by the easy and hard anisotropy axes for chosen orientations of the applied magnetic field \cite{cubic,droplets}. In contrast, skyrmions, due to their inherently two-dimensional structure, are more robust against such anisotropy-induced deformations, which enhances their stability.

In this work, we revisit the phenomenological Dzyaloshinskii framework \cite{Dz64} with the  incorporated cubic anisotropy \cite{Maleev} and systematically examine its influence on the magnetic phase diagram. Previous studies \cite{cubic,droplets} have largely overlooked the $(\text{cubic anisotropy})$–(magnetic field) phase space for negative anisotropy values. Here, we construct the corresponding phase diagram, determine the critical anisotropy threshold required to stabilize skyrmions, and delineate the regions in which the skyrmion lattice becomes energetically favorable compared to competing spiral states.

A central aspect of our analysis is the character of the transition between the conical spiral and the homogeneous state at the upper critical field $H_{c2}$. We demonstrate that, depending on the field orientation, this transition can be either first- or second-order, accompanied by rotation of the cone wave vector away from the field direction. Subtle reorientation processes of spiral vectors also occur near the lower critical field $H_{c1}$. Thus, our new theoretical results uncover previously unrecognized pathways by which cubic anisotropy reshapes the stability hierarchy of chiral modulated phases.  

Particular emphasis is placed on anisotropy-dependent parameters in the theoretical phase diagrams, which provide a direct connection between theory and experiment. Specifically, the angular dependence of the upper critical field $H_{c2}$ obtained from our calculations enables an estimate of the cubic anisotropy constant in the chiral magnet Fe$_{1-x}$Co$_x$Si.
Our results indicate that the cubic anisotropy in samples with $x = 0.08$ and $0.15$ is sufficiently strong to stabilize the low-temperature skyrmion lattice, potentially making Fe$_{1-x}$Co$_x$Si the first chiral metallic system, aside from the Mott-insulator Cu$_2$OSeO$_3$, to exhibit this phase.

Taken together, our findings establish cubic anisotropy as a decisive factor in stabilizing skyrmions far from $T_c$, extending the conventional picture of SkL formation confined to the high-temperature $A$-phase pocket. This framework provides a unified description of anisotropy-driven skyrmion stability and offers predictive guidelines for identifying and engineering LT-SkL phases in other chiral magnets.

\begin{figure*}[t]
  \centering
  \includegraphics[width=0.99\linewidth]{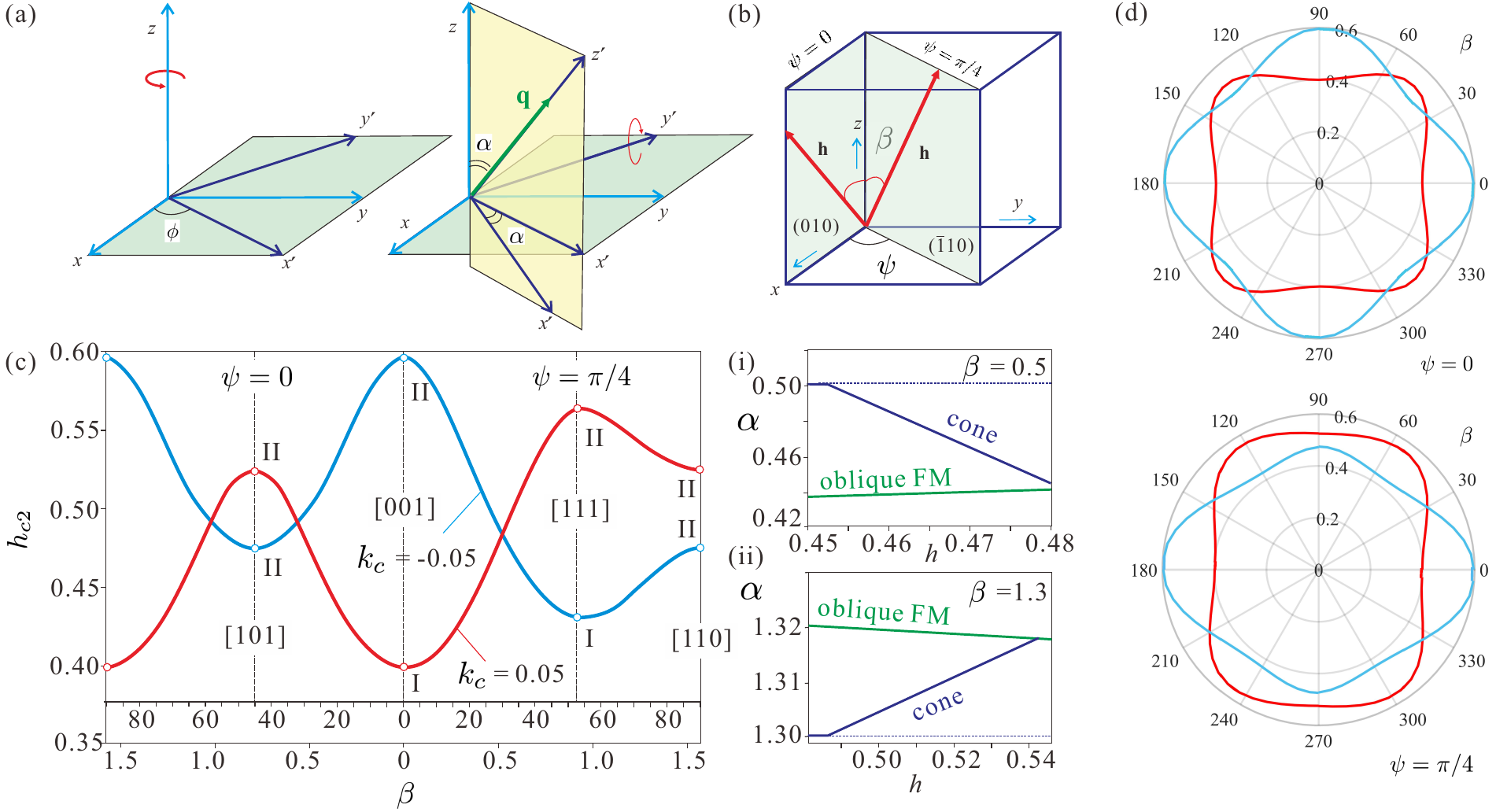}
  \caption{\label{fig01} (a) Construction of the rotated coordinate system $x^\prime y^\prime z^\prime$: the axes are obtained by a rotation through angle $\phi$ about the $z$ axis, followed by a rotation through angle $\alpha$ about the $y^\prime$ axis. The spiral wave vector is aligned with the $z^\prime$ axis. (b) Definition of the magnetic field orientation, described by the angle $\beta$, with the field confined to the $(010)$ plane for $\psi=0$ and to the $(\overline{1}10)$ plane for $\psi=\pi/4$. (c) Critical field $h_{c2}$ as a function of magnetic-field orientation within two selected crystallographic planes. First- and second-order transitions into the homogeneous state are indicated by I and II, respectively. Results are shown for $k_c=0.05$ (red curve) and $k_c=-0.05$ (blue curve). Insets (i) and (ii) illustrate the complex character of the transition from the conical state (dark-blue curves) to an oblique homogeneous state (green curves) for intermediate field directions $\beta=0.5$ and $\beta=1.3$. (d) Corresponding critical fields $h_{c2}$ in polar coordinates. 
 }
\end{figure*}

\begin{figure*}[t]
  \centering
  \includegraphics[width=0.99\linewidth]{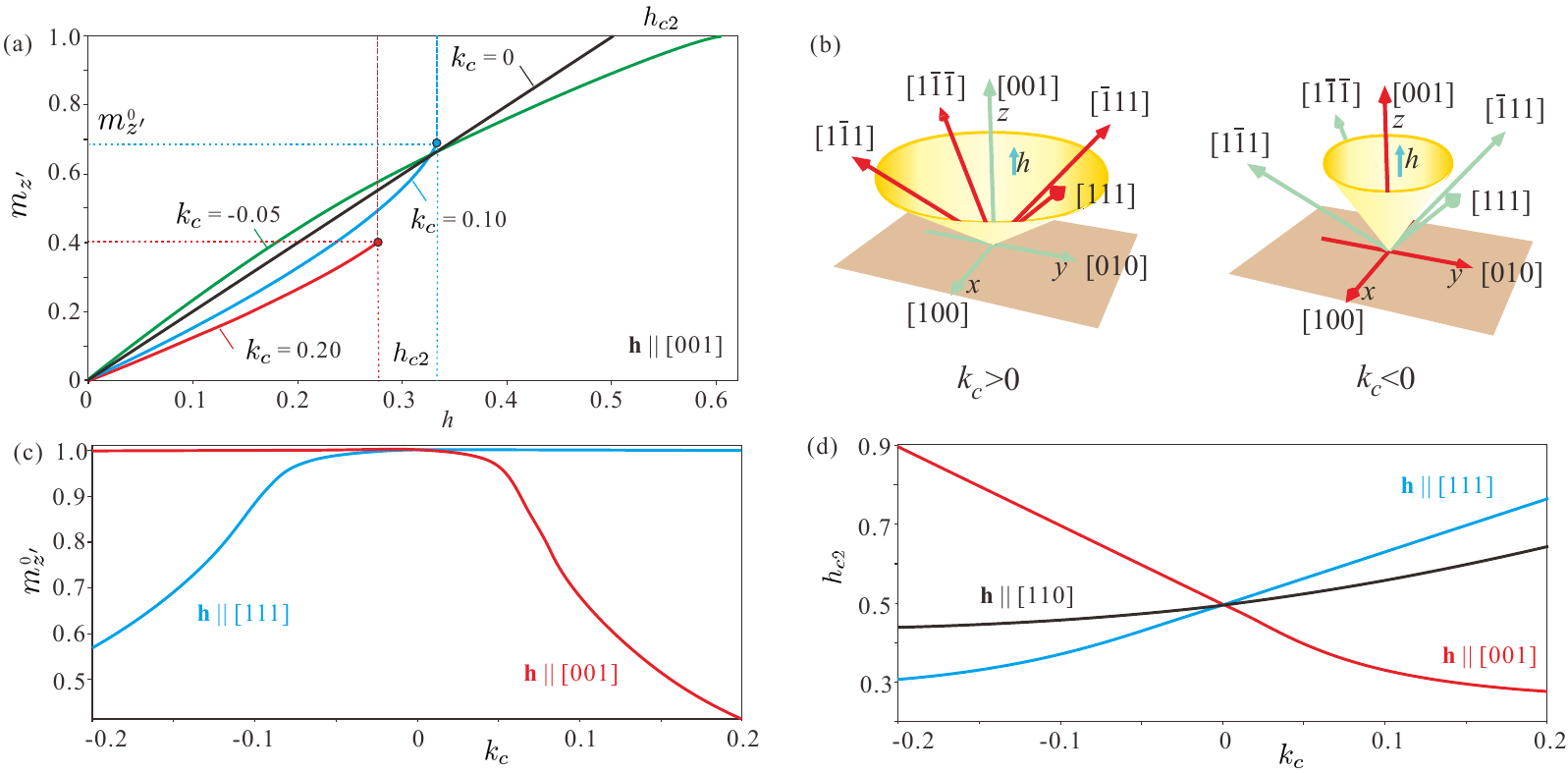}
  \caption{\label{fig02} 
Anisotropy-dependent magnetization behavior for different field orientations. 
(a) Magnetization curves for the field applied along $\mathbf{h}\parallel [001]$ for different values of $k_c$. For $k_c>0$, the curves exhibit jumps from the conical to the homogeneous state (red and blue curves), whereas for $k_c<0$, the magnetization increases smoothly up to saturation. (b) sketches of the magnetization rotation in the conical phase in the presence of cubic anisotropy with hard (red arrows) and easy (green arrows) axes for both signs of $k_c$.
(c) Maximal magnetization values immediately before the transition to the ferromagnetic state for $\mathbf{h} \parallel [111]$ and $\mathbf{h} \parallel [001]$. 
(d) Critical fields $h_{c2}$ along the three principal field directions. 
}
\end{figure*}

\section{Phenomenological model}

\subsection{Isotropic Dzyaloshinskii model}

Within the phenomenological framework introduced by Dzyaloshinskii \cite{Dz64}, the magnetic energy density of a bulk non-centrosymmetric ferromagnet with a spatially varying magnetization $\mathbf{M}$ is expressed as
\begin{equation}
w_0(\mathbf{M})=A \sum_{i,j}\left(\frac{\partial m_j}{\partial x_i}\right)^2
+D\,w_D(\mathbf{M}) -\mathbf{M}\cdot\mathbf{H}.
\label{density}
\end{equation}
Here, $A$ and $D$ denote the exchange and Dzyaloshinskii–Moriya interaction coefficients, respectively; $\mathbf{H}$ is the applied magnetic field, and $x_i$ are the Cartesian components of the spatial coordinate.

The DMI contribution $w_D$ is composed of Lifshitz invariants (LI)
\begin{equation}
\mathcal{L}^{(k)}_{i,j} = M_i \frac{\partial M_j}{\partial x_k} - M_j \frac{\partial M_i}{\partial x_k}
\label{LI}
\end{equation}
which are energy terms involving the first spatial derivatives of the magnetization.
For the purposes of this study, we perform all calculations assuming cubic helimagnets, where the Dzyaloshinskii–Moriya energy is given by  
\begin{equation}
w_{DMI} = \mathbf{m} \cdot \nabla \times \mathbf{m}
\label{DMI}
\end{equation}
Nevertheless, the formalism and results are applicable to magnets with other symmetry classes \cite{JETP89}, which may involve different combinations of Lifshitz invariants.

In chiral magnetism, the Dzyaloshinskii interaction functional (\ref{density}) plays a role analogous to the Frank functional in liquid crystals \cite{books} or the Ginzburg–Landau functionals in superconductivity \cite{Brandt95,Brandt03} and constitutes the primary theoretical framework for analyzing and interpreting experimental results in chiral magnets. Over several decades, systematic investigations of chiral modulations in diverse non-centrosymmetric magnetic systems have produced a substantial body of experimental data, which has been consistently interpreted within this approach (see, for example, the review \cite{Izyumov84} and the bibliography in Ref. \onlinecite{Nature06,roadmap}).

For the forthcoming calculations, we introduce dimensionless spatial coordinates $\mathbf{x} = \mathbf{r}/L_D$, where $L_D = A/D$ characterizes the periodicity of the modulated magnetic states. The unit vector along the magnetization direction is defined as 
$\mathbf{m}$, so that the magnetization  is $\mathbf{M} = M \mathbf{m}$. The applied magnetic field is similarly expressed in reduced units, $\mathbf{h} = \mathbf{H}/H_D$, with $\mu_0 H_D = D^2/(A M)$.

\subsection{Cubic anisotropy}
 
In cubic helimagnets such as Fe$_{1-x}$Co$_x$Si and MnSi, the isotropic energy density (\ref{density}) is supplemented by a cubic anisotropy contribution \cite{Bak80,Maleev},  
\begin{equation}
w_c = k_c \left( m_x^2 m_y^2 + m_x^2 m_z^2 + m_y^2 m_z^2 \right),
\label{anisotropies}
\end{equation}
where $k_c = K_c A / D^2$ denotes the reduced anisotropy constant, which in general depends on temperature. For typical cubic helimagnets, $K_c$ is found to be roughly one order of magnitude smaller than $A$, consistent with the hierarchy of magnetic interactions where cubic anisotropy is a relativistic correction to the isotropic exchange.

To generalize the analysis, we write the cubic anisotropy (\ref{anisotropies}) in a rotated coordinate system $x^\prime y^\prime z^\prime$, defined by the orientation of the spiral wave vector ($\mathbf{q}\parallel z^\prime$) and parameterized by the angles $\alpha$ and $\phi$ with respect to the original $xyz$ axes [Fig.~\ref{fig01}(a)]. The magnetization components in the two coordinate systems are related by:
\begin{align}
    m_x &= m_{x^\prime}\cos\alpha\cos\phi - m_{y^\prime}\sin\phi + m_{z^\prime}\sin\alpha\cos\phi, \nonumber\\
    m_y &= m_{x^\prime}\cos\alpha\sin\phi + m_{y^\prime}\cos\phi + m_{z^\prime}\sin\alpha\sin\phi, \nonumber\\
    m_z &= -m_{x^\prime}\sin\alpha + m_{z^\prime}\cos\alpha,
\end{align}
which enable the cubic anisotropy term (\ref{anisotropies}) to be expressed in the new coordinate system.

The magnetic field orientation with respect to the original $xyz$ axes is defined by the angles $\beta$ and $\psi$, such that  
$h_x = h \sin\beta \cos\psi$, $h_y = h \sin\beta \sin\psi$, and $h_z = h \cos\beta$ (Fig. \ref{fig01} (b)). Transforming to the rotated coordinate system, the field components become  
\begin{align}
    h_{x^\prime} &= h_x \cos\alpha \cos\phi + h_y \cos\alpha \sin\phi - h_z \sin\alpha, \nonumber\\
    h_{y^\prime} &= -h_x \sin\phi + h_y \cos\phi, \nonumber\\
    h_{z^\prime} &= h_x \sin\alpha \cos\phi + h_y \sin\alpha \sin\phi + h_z \cos\alpha,
\end{align}
enabling the Zeeman energy  
\begin{equation}
    w_{\mathrm{Zeeman}} = -h_{x^\prime} m_{x^\prime} - h_{y^\prime} m_{y^\prime} - h_{z^\prime} m_{z^\prime}
\end{equation}
to be expressed in the coordinate system associated with the spiral state.
For the simulations reported here, the magnetic field is restricted to lie within the $(010)$ plane for $\psi = 0$ and within the $(\overline{1}10)$ plane for $\psi = \pi/4$.

The isotropic exchange and Dzyaloshinskii–Moriya interactions retain the same functional form expressed via $m_{x^\prime}, m_{y^\prime},m_{z^\prime}$  components in the rotated coordinate system $x^\prime y^\prime z^\prime$.

We employ the MuMax3 software package (version 3.10) as the main numerical tool to minimize the functional (\ref{density}), which computes magnetization dynamics by solving the Landau–Lifshitz–Gilbert (LLG) equation using a finite-difference discretization approach \cite{mumax3}. To cross-check the reliability of the results, self-designed numerical routines are also used, as described in detail in \cite{metamorphoses}, and are therefore omitted here.

\section{Magnetic-Field-Driven Transformations of Spiral States in Helimagnets with Cubic Anisotropy}

\subsection{Magnetization curves}

The magnetization curves for different field directions provide direct information about the geometry of the easy and hard anisotropy axes, i.e., about the sign of $k_c$. 
For $k_c > 0$, the cubic anisotropy selects the $\langle 001 \rangle$ crystallographic axes as easy directions \cite{cubic}, while the $\langle 111 \rangle$ axes act as hard directions. In contrast, for $k_c < 0$, the $\langle 111 \rangle$ axes become easy.

The nature of the transition from the conical phase into the homogeneous phase at the critical field $h_{c2}$ has been thoroughly investigated in Refs.~\cite{cubic,droplets}. It was shown that for fields applied along the easy anisotropy axes, the cone closes via a first-order phase  transition (marked by ``I'' in Fig. \ref{fig01} (c)). In contrast, when the field is applied along the hard anisotropy axes, the transition is of second order (marked by ``II''). The phase transition for $\mathbf{h} \parallel \langle 110 \rangle$ is of second order \cite{cubic2} for both signs of $k_c$.

As an example, consider the case $k_c > 0$ with $\mathbf{h} \parallel [001]$ ($\beta=0$) (Fig.~\ref{fig02}(a)). Here, the magnetization in the conical phase is forced to rotate in the environment of hard $\langle 111 \rangle$ axes (Fig. \ref{fig02} (b)), which underlies the subsequent first-order transition into the homogeneous state. Consequently, the magnetization curves exhibit jumps from the conical to the homogeneous phase (red and blue curves in Fig. \ref{fig02} (a)). The size of this jump is directly related to the cubic anisotropy strength. 

For $k_c < 0$ with $\mathbf{h} \parallel [001]$ (Fig.~\ref{fig02}(a)), the $\langle 001 \rangle$ axes represent hard directions. In this case, the magnetization in the conical phase rotates to pass through the easy $\langle 111 \rangle$ directions (Fig. \ref{fig02} (b)). As a result, the transition remains second order, and the magnetization curve evolves smoothly up to saturation (green curve in Fig. \ref{fig02} (a)). 
Note the curvature change of the magnetization curves depending on the sign of $k_c$, and the straight-line behavior corresponding to the isotropic case ($k_c = 0$), given by $m_{z^\prime} = 2h$ (black curve), obtained from minimizing Eq.~(\ref{density}) for the conical state. 

For $\mathbf{h} \parallel [111]$ (not shown), the magnetization processes are qualitatively similar: a first-order transition occurs for $k_c < 0$, whereas a second-order transition -- for $k_c > 0$.

Figs.~\ref{fig02}(c),(d) summarize the anisotropy-dependent features obtained from the magnetization curves along easy and hard axes, and thus provide practical guidelines for estimating the value of $k_c$ from the characteristic signatures of the magnetization processes.
Fig.~\ref{fig02}(c) shows the maximal magnetization values just before the transition into the ferromagnetic state for two field directions, $\mathbf{h} \parallel [111]$ (blue curve) and $\mathbf{h} \parallel [001]$ (red curve). Finally, Fig.~\ref{fig02}(d) summarizes the critical fields $h_{c2}$ along the three principal field directions. The asymmetry of the graphs for different signs of the cubic anisotropy $k_c$ arises from the differing angles between the easy and hard axes. 

For the subsequent analysis of experimental magnetization curves, we introduce a more convenient measure of the non-dimensional anisotropy constant $k_c$, which simplifies its extraction from the experimental data.
As seen in Fig.~\ref{fig02}(a), the critical fields $h_{c2}$ vary with field orientation.  
This motivates the definition of the ratios
\begin{equation}
    \Delta_1 =
    \begin{cases}
        \dfrac{h_{c2[111]} - h_{c2[001]}}{h_{c2[001]}}, & k_c > 0, \\[2ex]
        \dfrac{h_{c2[001]} - h_{c2[111]}}{h_{c2[111]}}, & k_c < 0,
    \end{cases}
    \label{Delta1}
\end{equation}
which provide a practical means of quantifying $k_c$. In essence, the ratio $\Delta_1$ quantifies the difference between the maximal critical field $h_{c2}$ along the hard axis and the minimal field along the easy axis, normalized by the lowest critical field. It can be interpreted as a measure of the effectiveness of the cubic anisotropy.

\begin{figure*}
  \centering
  \includegraphics[width=0.9\linewidth]{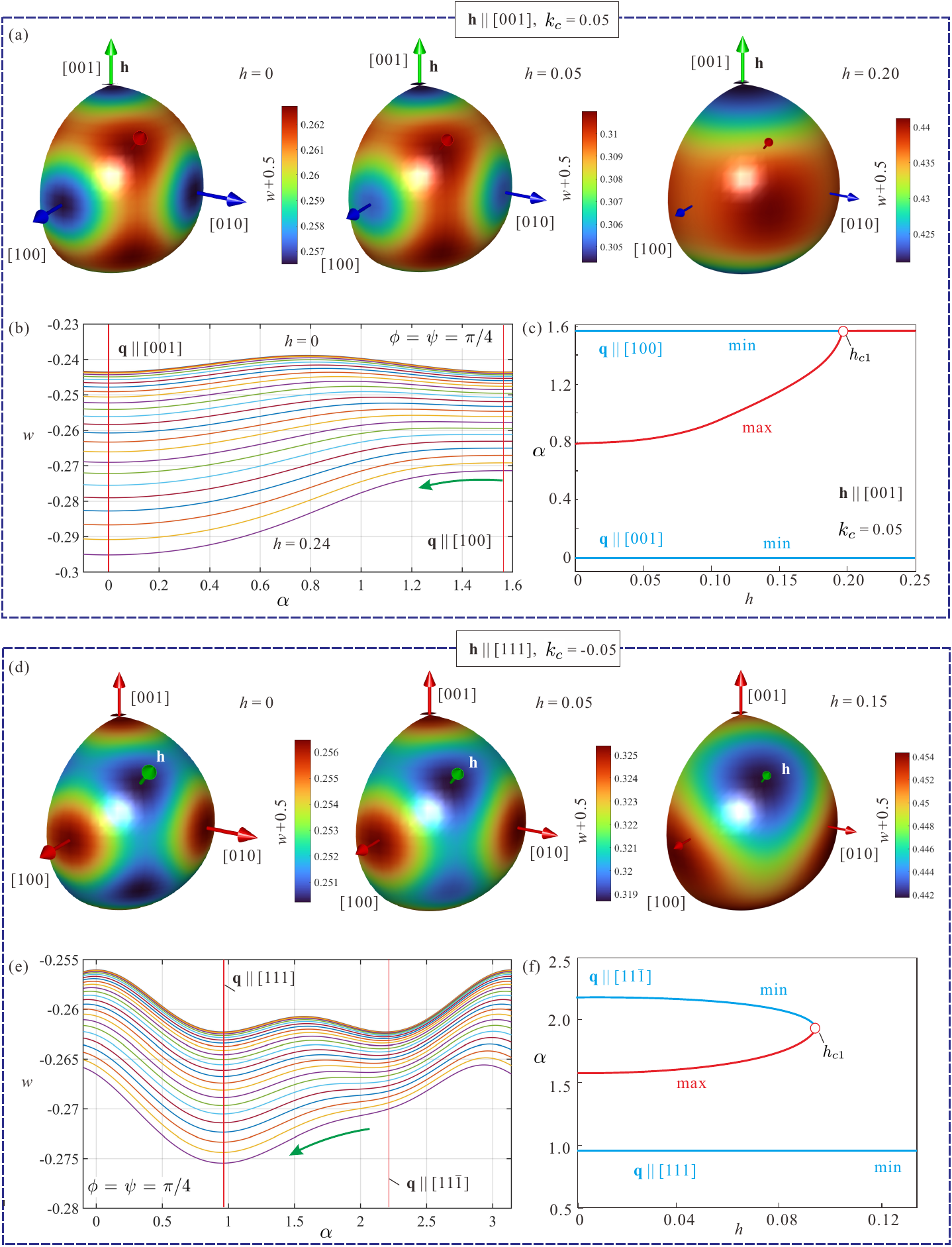}
  \caption{\label{fig03}  Reorientation of spiral states toward easy anisotropy axes at the critical field $h_{c1}$. 
(a) Energy landscape for $k_c > 0$ and $\mathbf{h} \parallel [001]$ in spherical coordinates, showing multidomain spiral states with mutually perpendicular wave vectors. 
(b) Energy density in the plane $\phi = \psi = \pi/4$ as a function of $\alpha$, used to identify the critical field $h_{c1}$ at which local minima corresponding to transverse spiral domains disappear. 
(c) Critical angles $\alpha$ corresponding to energy maxima (red curve) and minima (blue curves), with $h_{c1}$ indicated explicitly for $k_c = 0.05$. 
(d) Energy landscape for $k_c < 0$ and $\mathbf{h} \parallel [111]$, showing multidomain spiral states with energetically inequivalent $\langle 111 \rangle$ spirals. 
(e) Evolution of the energy landscape, illustrating the “jump” of metastable spirals into the global minimum. 
(f) Slight tilting of the spiral wave vectors toward the field direction for $k_c=-0.05$. 
}
\end{figure*}

 \begin{figure*}
  \centering
  \includegraphics[width=0.9\linewidth]{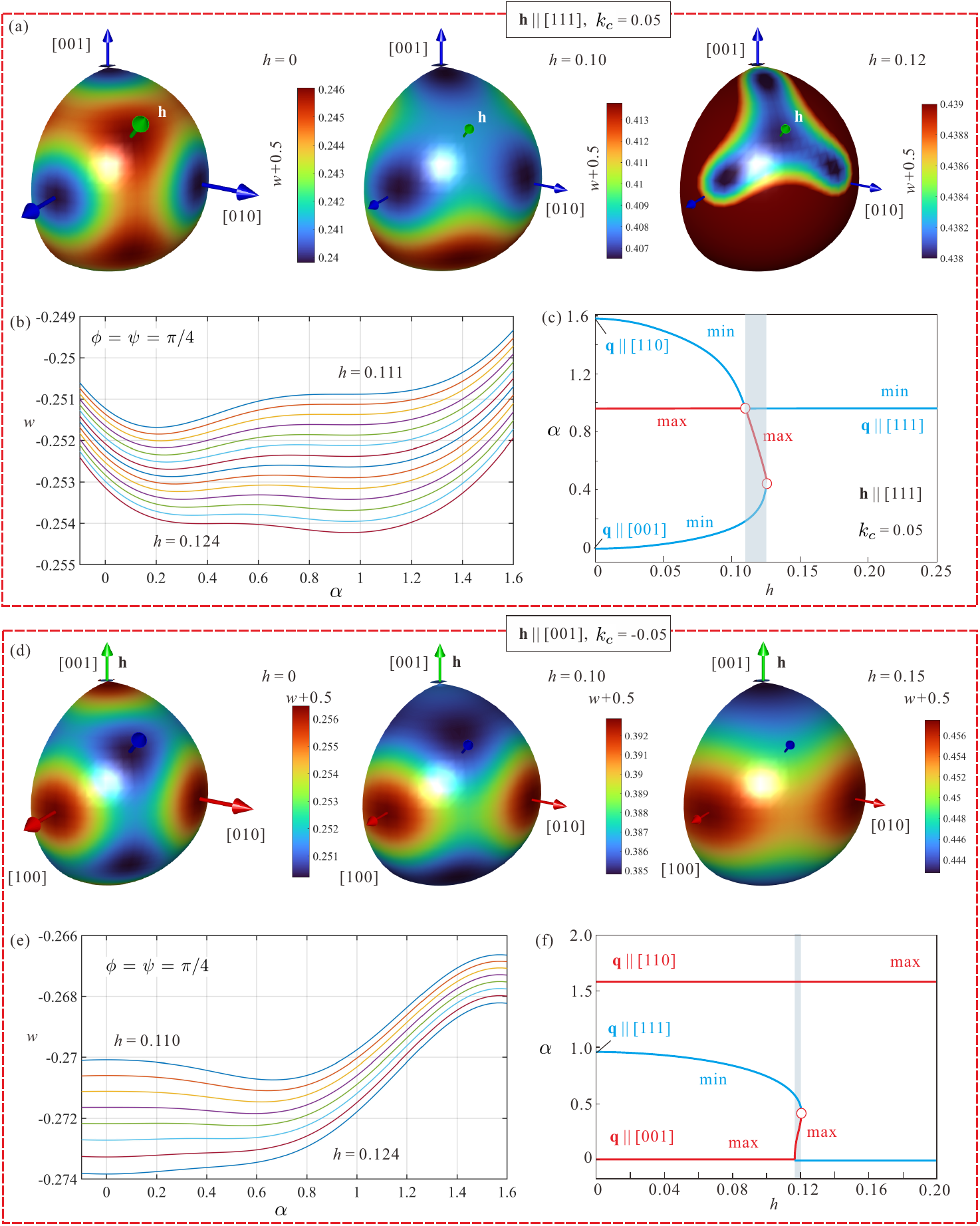}
  \caption{\label{fig04}  
Field-driven reorientation processes of spiral states along hard anisotropy axes.  
(a) Energy landscape of spiral states for $k_c > 0$ and $\mathbf{h} \parallel [111]$ shown in spherical coordinates, with the wave vectors $\mathbf{q} \parallel \langle 001 \rangle$ gradually canting towards the field.  
(b) 2D energy profiles indicating the emergence of a new minimum along the field direction at $h = 0.111$, its degeneracy with oblique spiral states at $h = 0.121$ (critical field $h_{c1}$), and the subsequent disappearance of the oblique minima at $h = 0.124$.  
(c) Field dependence of energy extrema illustrating the hysteretic region (shaded rectangle) where spiral and conical states coexist.  
(d–f) Analogous results for $k_c < 0$ and $\mathbf{h} \parallel [001]$, where a similar reorientation occurs but within a much narrower hysteretic interval: the conical minimum appears at $h = 0.116$, becomes degenerate with spiral minima at $h = 0.119$, and fully replaces them by $h = 0.120$.  
}
\end{figure*}

\subsection{The Conical-to-Homogeneous Phase Transition at the Critical Field $h_{c2}$}

Figure~\ref{fig01}(c) presents the critical fields $h_{c2}$ obtained from magnetic-field sweeps within the two characteristic planes $(010)$ and  $(\overline{1}10)$. In general, the critical curves exhibit three extrema along the characteristic crystallographic directions $\langle 001 \rangle$, $\langle 111 \rangle$, and $\langle 110 \rangle$. This observation allows us to define an additional field ratio
\begin{equation}
    \Delta_2 =
    \begin{cases}
        \dfrac{h_{c2[110]} - h_{c2[001]}}{h_{c2[001]}}, & k_c > 0, \\[2ex]
        \dfrac{h_{c2[110]} - h_{c2[111]}}{h_{c2[111]}}, & k_c < 0,
    \end{cases}
    \label{Delta2}
\end{equation}
which can be used to cross-check the previously defined ratio $\Delta_1$ for a given value of $k_c$.

To further elucidate the nature of the phase transitions for intermediate field orientations, we minimized the energy of the conical state with respect to the field direction for two representative values, $\beta=0.5$ and $\beta=1.3$ (see insets (i) and (ii) in Fig. \ref{fig01} (c)). Remarkably, although the spiral wave vector is already co-aligned with the applied field, it begins to deviate from this direction in the immediate vicinity of saturation. This deviation leads to the formation of an oblique homogeneous state, which subsequently rotates and aligns fully with the field, but for much larger values of the field. For realistic values of the anisotropy constant, however, such deviations remain rather small, occurring toward the $[001]$ axis for $\beta < 0.955$ and toward the $[110]$ axis for $\beta > 0.955$ (when $\mathbf{h}\parallel [111]$). The vanishingly small cone angle of the conical spiral during this process also presents certain difficulties for experimental identification.

\subsection{The helical-to-conical phase transition at the critical field $h_{c1}$}

In this section, we focus on the spiral flips occurring at $h_{c1}$, considering separately the cases where the field is applied along the easy and hard anisotropy axes.

\subsubsection{Spiral Reorientation toward Easy Anisotropy Axes}

Such reorientation processes occur for the field $\mathbf{h} \parallel \langle 001 \rangle$ when $k_c > 0$ and for $\mathbf{h} \parallel \langle 111 \rangle$ when $k_c < 0$ (Fig.~\ref{fig03}). In both cases, the spiral domain with its wave vector aligned along the field immediately acquires an energetic advantage over other domains with perpendicular ($k_c > 0$) or canted ($k_c < 0$) orientations of their wave vectors relative to the field. In other words, the conical phase persists throughout the entire field range up to saturation.

Figs.~\ref{fig03}(a) and (d) show the energy density of spiral states in spherical coordinates as a function of their orientation relative to the magnetic field, i.e.:
\begin{equation}
    w = \frac{1}{p} \int_0^p \left[ w_0 + w_c \right] dz^\prime,
\end{equation}
Here, the anisotropy energy density \(w_c\) and the Zeeman energy are expressed in the rotated coordinate system \(x^\prime y^\prime z^\prime\).
\(p\) is the equilibrium spiral period, which is close to the value \(4\pi L_D\) for the considered magnetization processes.
The radii of the spheres represent the energy values. Since the energy density is negative, we rescale it by a factor of 0.5 to make it positive and further subtract the energy of the homogeneous state co-aligned with the field.

\begin{figure}[t]
  \centering
  \includegraphics[width=0.99\linewidth]{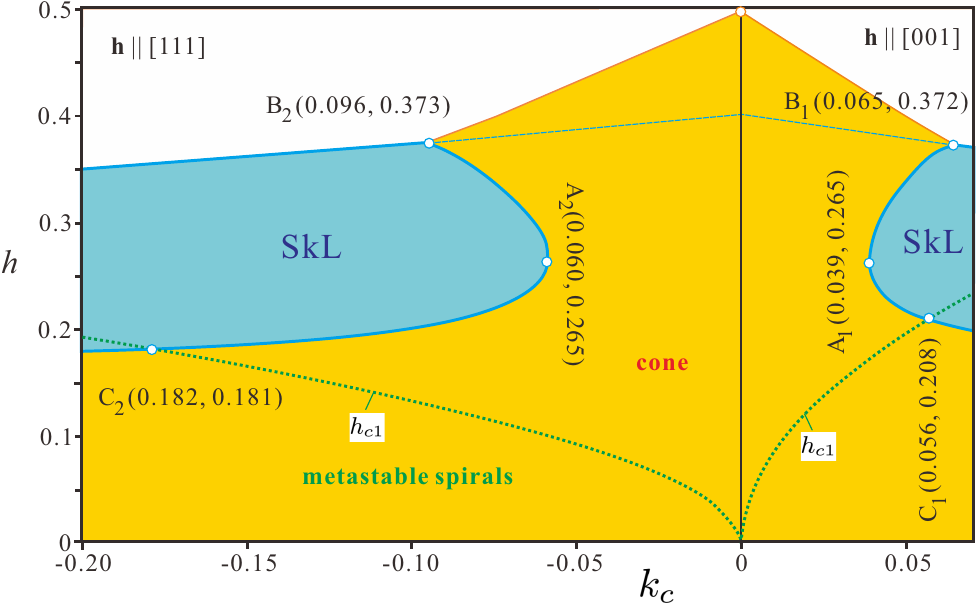}
  \caption{\label{fig05} 
Phase diagrams showing the stability regions of the low-temperature skyrmion lattice in cubic helimagnets.  
For $k_c>0$ and $\mathbf{h} \parallel [001]$, the vast area of SkL stability is highlighted. Critical points $A_1$, $B_1$, and $C_1$ indicate, respectively, the onset of the SkL, the threshold for the second-order transition to the homogeneous state, and the anisotropy value above which metastable spirals can act as nucleation centers for skyrmions.  
For $k_c<0$ and $\mathbf{h} \parallel [111]$, the SkL stability exhibits qualitatively similar features, with critical points $A_2$ and $C_2$ corresponding to the onset of the SkL and the nucleation threshold via metastable spirals.}
\end{figure}

For $k_c > 0$ and $\mathbf{h} \parallel [001]$, the multidomain spiral state consists of spiral domains with mutually perpendicular wave vectors, $\mathbf{q} \parallel \langle 001 \rangle$ (Fig.~\ref{fig03}(a)). As the field increases, the minima corresponding to the [100] and [010] domains eventually vanish, marking the critical field $h_{c1}$, after which only the domain of the conical phase remains (see also supplementary videos). To identify the field at which a local minimum disappears, we plot the energy density in the plane defined by $\phi = \psi = \pi/4$ while varying the angle $\alpha$ (Fig.~\ref{fig03}(b)). Additionally, Fig.~\ref{fig03}(c) shows the critical spiral orientations corresponding to energy maxima (red curve) and energy minima (blue curve), with $h_{c1}$ explicitly indicated for $k_c = 0.05$. This figure also shows that the wave vectors of the spiral states remain strictly perpendicular to the field throughout the entire field range below $h_{c1}$.
The anisotropy-dependent values of $h_{c1}$ are summarized in the phase diagram shown in Fig.~\ref{fig05}.
According to theoretical predictions, the domains of transverse spirals do not reappear upon lowering the field, consistent with the experimental observations reported in Ref.~\onlinecite{Halder}.

For $k_c < 0$ and $\mathbf{h} \parallel [111]$, the multidomain spiral state consists of energetically inequivalent $\langle 111 \rangle$ spirals (Fig.~\ref{fig03}(d)). Fig.~\ref{fig03}(e) illustrates the corresponding process in which the local minima associated with metastable spirals eventually "jump" into the global minimum of the conical state once the energy barrier vanishes. Fig.~\ref{fig03}(f), however, shows that the spiral wave vectors slightly tilt toward the field. The anisotropy-dependent values of $h_{c1}$ for this field direction are also summarized in the phase diagram of Fig.~\ref{fig05}.

\subsubsection{Spiral Reorientation toward Hard Anisotropy Axes}
For field directions along the hard anisotropy axes, all spiral domains correspond to the global energy minima of the functional (\ref{density}). At low fields, no conical state aligned with the field exists; such a conical phase only emerges in the vicinity of $h_{c1}$.

For $k_c > 0$ and $\mathbf{h} \parallel [111]$, spiral wave vectors $\mathbf{q} \parallel \langle 001 \rangle$ gradually cant towards the field (Fig.~\ref{fig04}(a)). The 2D energy map in Fig.~\ref{fig04}(b) shows that an additional energy minimum along the field direction $[111]$ develops at $h = 0.111$, coexisting with the minimum of oblique spiral states. At $h = 0.121$, the two minima become degenerate, identifying the critical field $h_{c1}$. At slightly higher fields ($h = 0.124$), the original spiral minima vanish, and only the conical phase persists. This field interval is highlighted by the shaded rectangle in Fig.~\ref{fig04}(c). We note that the labels “min” and “max” refer to features in the 2D crosscuts of the energy landscape; as seen from the full 3D plots in Fig.~\ref{fig04}(a), spirals with $\mathbf{q} \parallel [110]$ correspond to points lying on the energy slope rather than genuine extrema.

For $k_c < 0$ and $\mathbf{h} \parallel [001]$, a qualitatively similar reorientation process occurs, although the associated hysteresis loop is much narrower (Fig. \ref{fig04} (d)-(f)). The minimum corresponding to the conical phase first appears at $h = 0.116$ and becomes degenerate with the original spiral minima at $h = 0.119$. Shortly thereafter, at $h = 0.120$, the oblique spiral states vanish entirely, leaving only the conical phase.

\begin{figure}[t]
  \centering
  \includegraphics[width=0.92\linewidth]{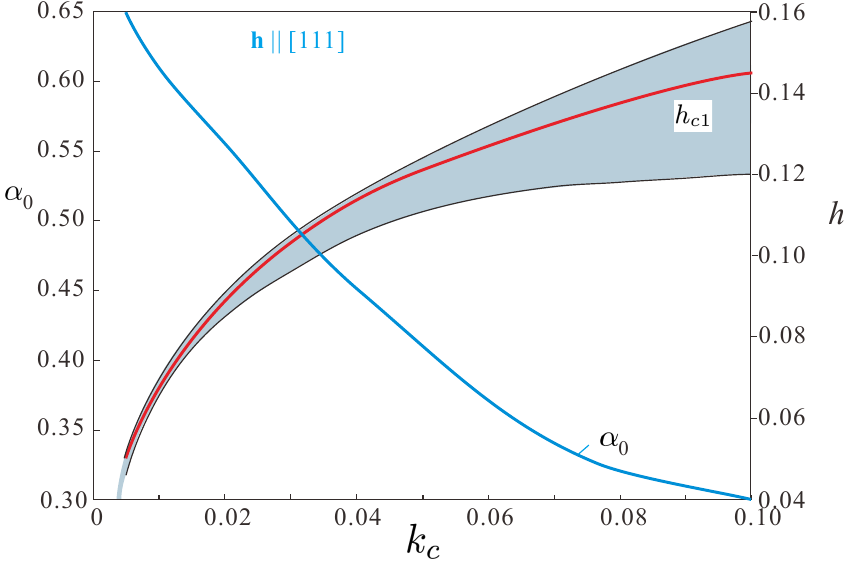}
  \caption{\label{fig06} Critical fields governing the hysteretic spiral reorientation for $\mathbf{h}\!\parallel\![111]$ with $k_c>0$ (right scale). The red curve indicates $h_{c1}$, while in the shaded region two minima coexist, corresponding to the oblique spiral and the conical state. The blue curve (left scale) shows the maximum oblique angle of metastable spirals prior to their abrupt alignment with the field.}
\end{figure}

We summarize the anisotropy-dependent critical fields in Fig.~\ref{fig06} for the case $k_c > 0$ and $\mathbf{h} \parallel [111]$, while omitting the counterpart for $k_c < 0$ and $\mathbf{h} \parallel [001]$ due to the narrowness of the corresponding field range. The ``magic'' angle $\alpha_0$ marks the threshold at which metastable spirals eventually flip into alignment with the field at the upper boundary of the histeretic region. It approaches $\alpha_0 = 0.955$ in the limit $k_c \to 0$, $h_{c1} \to 0$ (Fig.~\ref{fig06}).  

\subsubsection{Spiral reorientations for arbitrary field directions}

For arbitrary field orientations with respect to the easy and hard anisotropy axes, a single field cannot be assigned to $h_{c1}$. Instead, the first critical field corresponds to the transition of the spiral whose wave vector is closest to the field direction, while subsequent jumps mark the critical fields for more distant wave vectors. Thus, in general, there are three distinct $h_{c1}$ fields for $k_c>0$ and four for $k_c<0$. 

As an example, for $\mathbf{h}\!\parallel\![110]$ with $k_c>0$, the spirals with $\mathbf{q}\!\parallel\![100]$ and $\mathbf{q}\!\parallel\![010]$ flip first along the field, while the reorientation of the metastable spiral with $\mathbf{q}\!\parallel\![001]$ occurs at a larger field. A slight displacement of the field from the $[110]$ direction, e.g., with $\psi=0.6, \beta=\pi/2$, lifts the degeneracy so that the $\mathbf{q}\!\parallel\![100]$ and $\mathbf{q}\!\parallel\![010]$ spirals flip at different $h_{c1}$.  

Since determining all such critical fields for arbitrary field orientations requires computationally intensive numerical simulations, we do not pursue them here.

\section{Phase diagrams of states with stable LT-SkLs}

The phase diagram for $k_c > 0$ and $\mathbf{h} \parallel [001]$, (Fig. \ref{fig05}) which exhibits the large area of SkL stability (blue shading), has been described in several of our previous works \cite{cubic,cubic2,droplets}. 
Here, we briefly summarize the significance of some critical points: 
\begin{itemize}
  \item $k_c(A_1) = 0.039$ marks the smallest anisotropy value at which the SkL first appears;
  \item $k_c(B_1) = 0.065$ indicates the threshold beyond which the SkL undergoes a second-order phase transition into the homogeneous state at the upper boundary of its stability region. Below this value, the SkL transforms via a first-order phase transition into the conical state (yellow shading);
  \item $k_c(C_1) = 0.056$ marks the threshold where metastable spirals below $h_{c1}$ can serve as nucleation media for skyrmions: the spirals may transform into the thermodynamically stable SkL through the formation of merons, i.e., ruptures of a spiral state each carrying a topological charge $Q = 1/2$ \cite{Mukai}.
\end{itemize}

The portion of the phase diagram for $k_c < 0$ and $\mathbf{h} \parallel [111]$ qualitatively exhibits the same features, although all critical points take different values. The onset of the SkL occurs for $k_c > k_c(A_2) = 0.060$, which is significantly larger than in the previous case. Metastable spiral domains can act as nucleation centers for skyrmions only for $k_c > k_c(C_2) = 0.182$. 

\begin{figure*}[t]
  \centering
  \includegraphics[width=0.99\linewidth]{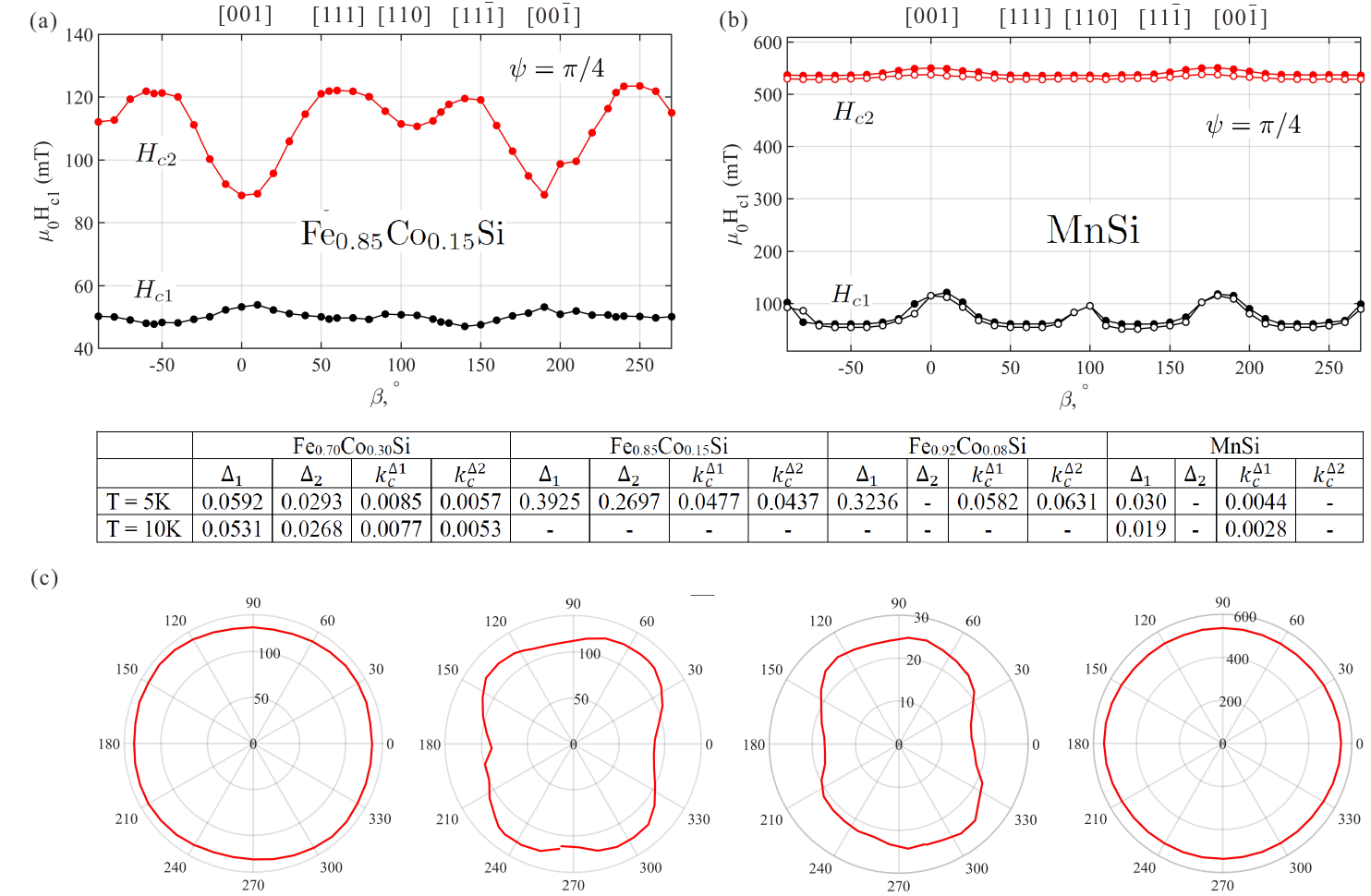}
  \caption{\label{fig07} 
  Critical fields $H_{c1}$ (black curves) and $H_{c2}$ (red curves) as functions of crystal orientation at $T = 5\,\text{K}$ for Fe$_{0.85}$Co$_{0.15}$Si (a), and at $T = 5\,\text{K}$ and $10\,\text{K}$ for MnSi (b). 
  The values of $H_{c1}$ and $H_{c2}$ have been corrected for the projected demagnetization field. 
  The table summarizes the ratios $\Delta_{1,2}$ from (\ref{Delta1}) and (\ref{Delta2}) and the corresponding non-dimensional anisotropy constants $k_{c}^{\Delta_1}$ and $k_{c}^{\Delta_2}$. (c) The corresponding polar plots $H_{c2}$.}
\end{figure*}

\section{Experimental results}

The theoretical results presented in the previous sections provide the foundation for a quantitative comparison with experimental data for specific chiral magnets. Eventually, our aim is to extract the value of the non-dimensional cubic anisotropy constant $k_c$ and thereby locate the experimentally observed magnetization processes on the theoretical phase diagram shown in Fig.~\ref{fig05}.

\subsection{Angular dependencies of the critical fields}
For the experimental study of the angular dependence of the critical fields, single crystals of MnSi and Fe$_{1-x}$Co$_{x}$Si ($0.08 \leq x \leq 0.70$) were grown using the tri-arc Czochralski method, and magnetization measurements are taken in a commercial SQUID setup as described in detail in Ref. \cite{next}. Representative data for Fe$_{0.85}$Co$_{0.15}$Si at $T = 5\,\text{K}$ with $\psi = 45^\circ$ are shown in Fig.~\ref{fig07}(a). The corresponding data for MnSi at $T = 5\,\text{K}$ (solid markers) and $T = 10\,\text{K}$ (open markers) are shown in Fig.~\ref{fig07}(b). 

The angular dependence of the experimental $H_{c2}$ curves closely matches the theoretical predictions shown in Fig.~\ref{fig01}(c). For Fe$_{1-x}$Co$_{x}$Si at low Co concentrations ($x \leq 0.20$), the critical field $H_{c2}$ reaches its maximum along the hard axes and exhibits a minimum along the easy axes, whereas $H_{c1}$ displays the opposite trend.  In MnSi, the critical fields follow the same angular dependence (Fig.~\ref{fig01}(c)), but the sequence of maxima and minima is reversed compared to Fe$_{1-x}$Co$_{x}$Si. Notice, however, that for arbitrary field orientations, as discussed in Sect.~III.C.3, the values of $H_{c1}$ correspond to averages over multiple spiral reorientation events. In general, the first-order phase transitions at $H_{c1}$ are expected to display hysteresis and domains of coexisting phases, which tend to smooth the jumps of the magnetization in the observed magnetization curves.

\begin{figure}[t]
  \centering
  \includegraphics[width=0.99\linewidth]{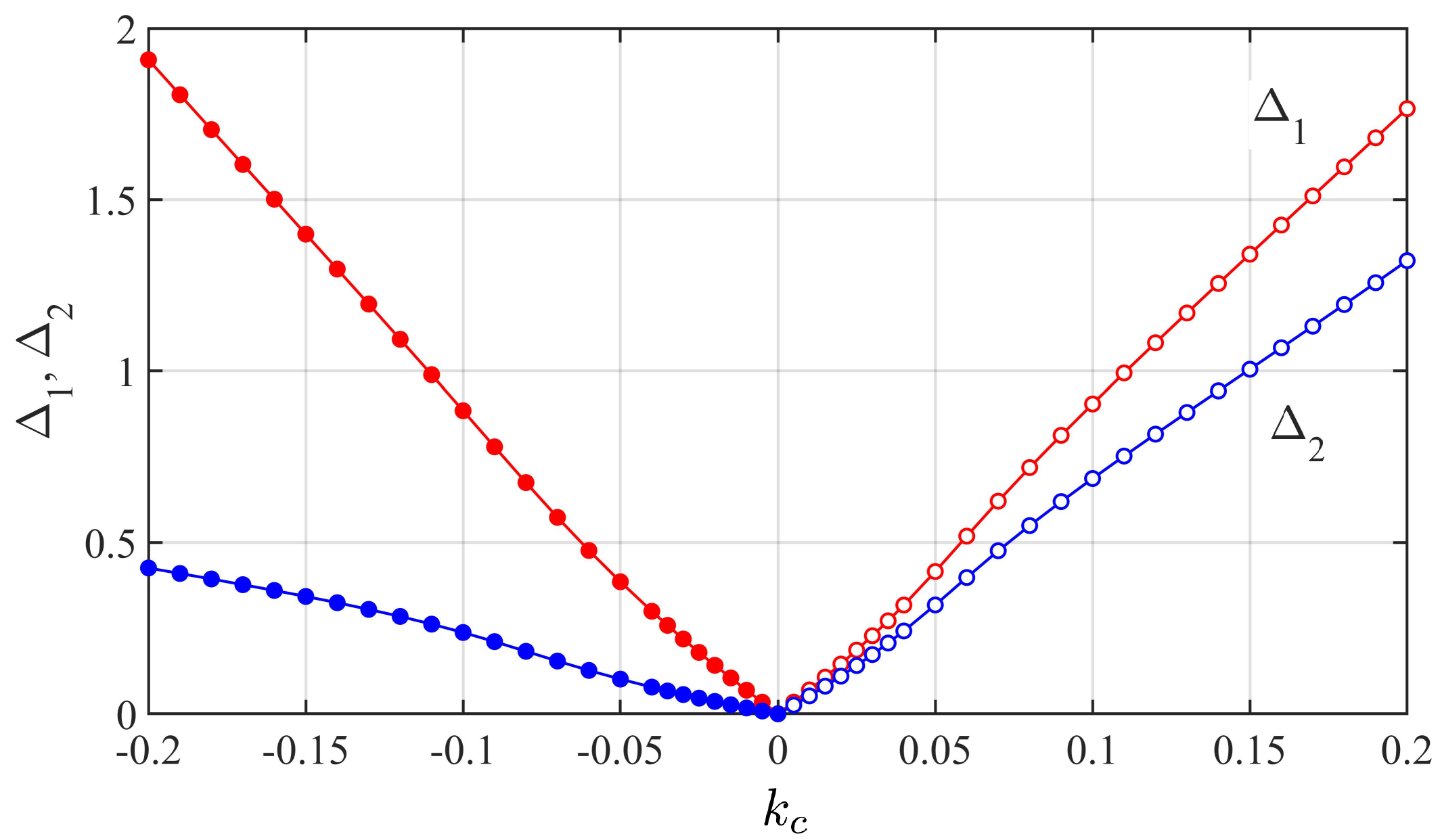}
  \caption{\label{fig08} 
  Anisotropy-dependent parameters $\Delta_1$ and $\Delta_2$ as defined in (\ref{Delta1}) and (\ref{Delta2}) for the critical fields along the three principal directions $[001]$, $[110]$, and $[111]$ and both signs of the anisotropy coefficient $k_c$. 
  }
\end{figure}

The experimentally determined critical fields along the three principal crystallographic directions enable the evaluation of the ratios $\Delta_1$~(\ref{Delta1}) and $\Delta_2$~(\ref{Delta2}), from which 
values of $k_c$ can be obtained by comparison with theory. Figure~\ref{fig08} displays the theoretically computed ratios $\Delta_1$ and $\Delta_2$ based on the critical field data shown in Fig.~\ref{fig02}(d). The table in Fig.~\ref{fig07} summarizes the anisotropy constants $k_{c}^{\Delta_1}$ and $k_{c}^{\Delta_2}$ extracted from $\Delta_1$ and $\Delta_2$, respectively, using the available experimental data at $T=5$ K and $10$ K.

The extracted values of $k_{c}^{\Delta_1}$ and $k_{c}^{\Delta_2}$ are found to be mutually consistent within the experimental uncertainty. For quantitative analysis, however, the ratio $\Delta_1$ is generally more reliable, as it exhibits a larger contrast between the corresponding critical fields. 

\subsection{Analysis of the experimental results}

\subsubsection{MnSi and Fe$_{0.70}$Co$_{0.30}$Si.}

As summarized in the table of Fig.~\ref{fig07}, both Fe$_{0.70}$Co$_{0.30}$Si and MnSi display cubic anisotropy values 
approximately an order of magnitude smaller than the threshold required to stabilize the LT-SkL. 
In these cases, the polar plots [Fig.~\ref{fig07}(c)] are nearly indistinguishable from perfect circles.
For MnSi, the evaluation of $k_{c}^{\Delta_2}$ is not feasible, since the difference between the critical fields along $[110]$ and $[111]$ lies within the experimental error. In both compounds, the extracted anisotropy constants decrease with increasing temperature, indicating a clear temperature dependence. 

\subsubsection{Fe$_{0.85}$Co$_{0.15}$Si.}

In Fe$_{0.85}$Co$_{0.15}$Si, the estimated cubic anisotropy $k_c^{\Delta_1} \approx 0.0477$ exceeds the critical threshold $k_c(A_1) = 0.039$, indicating that skyrmion stabilization is possible, though restricted to a narrow stability range. As shown in Fig.~\ref{fig05}, the corresponding lower and upper field boundaries of the SkL pocket are located at $h = 0.221$ and $h = 0.330$, respectively. 

This imposes specific constraints on skyrmion nucleation:
\begin{itemize}
    \item First, since the cubic anisotropy constant remains below the critical point $C_1$ (Fig.~\ref{fig05}), spiral states cannot efficiently serve as nucleation pathways for skyrmions.
    \item Second, the formation of a SkL directly from the conical phase by sweeping the magnetic field is also hindered. Because the cubic anisotropy constant remains below the critical value $k_c(B_1) = 0.065$, nucleation of isolated skyrmions \cite{torons} would occur only within the conical phase rather than the homogeneous state. This leads to an energetically unfavorable mismatch between isolated skyrmions and the conical background, so stability is attained only when they condense into a lattice.
\end{itemize}

 In this situation, skyrmions are more likely to nucleate upon cooling from the paramagnetic phase, passing through the A-phase. This, however, raises the question of how to distinguish “supercooled” metastable skyrmions from genuinely stable states. Another possible route would be first to estimate $k_c$ at lower temperatures: indeed, the value $k_c = 0.0477$ at $5\,\text{K}$ in Fe$_{85}$Co$_{15}$Si lies quite close to the critical value $k_c(C_1) = 0.056$, suggesting that the threshold might indeed be reached under suitable conditions.

As a consistency check, we also compare the critical fields $h_{c1}$ obtained theoretically with the experimental values. For Fe$_{0.85}$Co$_{0.15}$Si, the theoretical prediction for $\mathbf{h}\parallel [001]$ is $h_{c1}=0.191$, corresponding to the ratio $h_{c1}/h_{c2}=0.191/0.402=0.475$. Experimentally, we find $H_{c1}=53.9\,\text{mT}$, yielding $H_{c1}/H_{c2}=53.9/88.7=0.608$. The fact that the experimental ratio is slightly larger suggests that the reorientation from metastable spiral states into the conical phase is triggered only once the corresponding energy minimum vanishes. In addition, pinning effects slow down the relaxation processes and are expected to further shift $H_{c1}$ to higher values. Interestingly, measurements performed with slower magnetization sweeps in Ref.~\cite{next} yielded a somewhat reduced $H_{c1}$, bringing the experimental ratio into near-perfect agreement with the theoretical prediction.

For $\mathbf{h}\parallel [111]$ in Fe$_{0.85}$Co$_{0.15}$Si, the theoretical predictions are $h_{c1}=0.119$ and $h_{c1}/h_{c2}=0.119/0.560=0.213$. 
Experimentally, we obtain $H_{c1}=47.03\,\text{mT}$, corresponding to $H_{c1}/H_{c2}=47.03/124=0.38$. 
As in the [001] case, the experimental ratio is noticeably larger than the theoretical one, which we attribute to the same effects—namely, the persistence of metastable states, pinning phenomena, and slow relaxation processes that effectively shift $H_{c1}$ to higher values.

\subsubsection{Fe$_{0.92}$Co$_{0.08}$Si.}

In Fe$_{0.92}$Co$_{0.08}$Si, the cubic anisotropy $k_c^{\Delta_1} \approx 0.0582$ exceeds both the critical threshold $k_c = 0.039$, indicating the possibility of skyrmion stabilization, and the value $k_c(C_1)$, which implies an active role of metastable spiral states in skyrmion nucleation.

The value $k_c^{\Delta_2} \approx 0.0631$ further suggests that the threshold $k_c(B_1)$ may be surpassed upon lowering the temperature, thereby opening a nucleation pathway for skyrmions through torons and bobbers \cite{torons}.

\section{Conclusions}

In this work, we revisited the phenomenological Dzyaloshinskii framework as the primary theoretical tool for analyzing magnetization processes in bulk chiral magnets and systematically investigated the role of cubic magnetocrystalline anisotropy. We constructed phase diagrams for various field orientations and identified the characteristic signatures of cubic anisotropy in the evolution of helical and conical states, as well as in the stabilization of low-temperature skyrmion lattices.

We introduced convenient dimensionless measures, $\Delta_1$ and $\Delta_2$, which characterize the differences in $H_{c2}$ fields along principal crystallographic directions. This enabled the extraction of the non-dimensional cubic anisotropy constant $k_c=K_cA/D^2$ from experimental magnetization curves. The theoretical analysis demonstrated that the ratios $\Delta_{1,2}$ provide consistent and reliable estimates of $k_c$, and highlighted the critical threshold, $k_c \approx 0.039$, required to stabilize LT-SkL phases.

Angle-resolved magnetization measurements on MnSi and Fe$_{1-x}$Co$_x$Si single crystals confirmed the theoretical predictions. Both MnSi and Fe$_{0.70}$Co$_{0.30}$Si exhibit cubic anisotropy values close to zero, roughly an order of magnitude below the threshold required for LT-SkL stabilization. In Fe$_{0.85}$Co$_{0.15}$Si, $k_c$ slightly exceeds the critical value, suggesting that skyrmion stability is possible within a narrow field range. The largest anisotropy was observed in Fe$_{0.92}$Co$_{0.08}$Si, where $k_c$ significantly surpasses the threshold, providing a robust region of skyrmion stability. In all samples, $k_c$ exhibits a clear temperature dependence, decreasing with increasing temperature. Overall, in contrast to MnSi, the distinctive feature of the Fe$_{1-x}$Co$_{x}$Si series is that key parameters—including the ordering temperature, helical modulation length, spin-wave stiffness, Dzyaloshinskii–Moriya interaction strength, and cubic anisotropy—can be systematically tuned by varying the Co concentration, as has been demonstrated in a wide range of experiments \cite{Beille_1981,Manyala2004,Grefe2024}.

Although torque magnetometry \cite{Sauther} can provide more accurate measurements of cubic anisotropy, it is not always readily available. Similarly, small-angle neutron scattering (SANS) can independently determine anisotropy-related parameters introduced in this work, such as the critical angles $\alpha_0$ associated with spiral reorientation at $H_{c1}$, and thus serve as a cross-check for our estimates of $k_c$. In contrast, the magnetization-based approach presented here is widely accessible and offers a convenient  method for evaluating $k_c$ across a broad range of samples.

Thus, the combination of theoretical modeling and experimental measurements establishes cubic anisotropy as a tunable control parameter for skyrmion stability in itinerant chiral magnets. This methodology can guide the identification of candidate materials for LT-SkL phases, inform experimental designs probing anisotropy-driven stabilization, and facilitate the interpretation of magnetization data in terms of underlying magnetocrystalline anisotropy. Future work may combine magnetization-derived measurements with torque magnetometry and SANS experiments to achieve consistent determinations of \(K_c\), alongside further refinement of theoretical models to capture the full complexity of spin textures.

\end{document}